\documentstyle[11pt,mrs2003,epsfig]{article}
\begin{document} 

\newcommand{\sgn}{\mathop{\mathrm{sgn}}}
\newcommand{\md}{{\mathrm{d}}}

\mbox{}\hfill \parbox[t]{3.5cm}{CGPG--03/9--2\\ AEI--2003--077}

\title{QUANTUM GRAVITY AND THE BIG BANG}

\author{MARTIN BOJOWALD}
\affil{Center for Gravitational Physics and Geometry, The Pennsylvania
State University, 104 Davey Lab, University Park, PA 16802, USA}

\affil{New Address: Max-Planck-Institut f\"ur
Gravitationsphysik, Albert-Einstein-Institut, Am M\"uhlenberg 1,
D-14476 Golm, Germany; e-mail: {\tt mabo@aei.mpg.de}}

\begin{abstract} 
 Quantum gravity has matured over the last decade to a theory which can
tell in a precise and explicit way how cosmological singularities of
general relativity are removed. A branch of the universe ``before'' the
classical big bang is obtained which is connected to ours by quantum
evolution through a region around the singularity where the classical
space-time dissolves. We discuss the basic mechanism as well as
applications ranging to new phenomenological scenarios of the early
universe expansion, such as an inflationary period.
\end{abstract} 
 
\section{Introduction} 

 When the big bang is approached, the volume becomes smaller and
smaller and one enters a regime of large
 energy
densities. Classically, those conditions will become so severe
 that a
singularity is reached; the theory simply breaks down. For a
 long
time, the expectation has been that somewhere along the way
 quantum
gravity takes over and introduces new effects, e.g.\ a
 discrete
structure, which prevent the singularity to develop. This
 presumably
happens at scales the size of the Planck length $\ell_P$,
 i.e.\ when
the universe has about a volume $\ell_P^3$.
 
 Since at the classical
singularity space itself becomes singular and
 gravitational
interactions are huge, such a quantum theory of gravity
 must be
background independent and non-perturbative. A theory satisfying these
conditions is
 in fact available in the form of loop quantum
gravity/quantum geometry
 (see
\cite{Rov:Loops,ThomasRev} for reviews). One of its early
 successes
was the derivation of discrete
 spectra of geometric operators like
area and volume
\cite{AreaVol,Area,Vol2}. Thus, the spatial geometry is discrete in a
precise sense. Furthermore, matter Hamiltonians exist as well-defined
operators in the theory which implies that ultraviolet
divergences are cured in the fundamental formulation \cite{AnoFree,QSDI}.

Both properties must be expected to have important consequences for
cosmology. The discreteness leads to a new basic formulation
valid at small volume, and since gravity couples to the matter
Hamiltonian, its source term is modified at small scales when the good
ultraviolet behavior is taken into account. It is
possible to introduce both effects into a cosmological model in a
systematic way, which allows us to test the
cosmological consequences of quantum gravity (reviewed in \cite{Essay,
LoopCosRev}).

\section{Cosmological evolution equations} 

Classically, the dynamics of a flat isotropic universe is described by
the Friedmann equation
\begin{equation} \label{Friedmann}
 \left(\frac{\dot{a}}{a}\right)^2= \frac{16\pi G}{3}\rho(a)
\end{equation}
where we can choose the energy density of a single scalar,
\begin{equation} \label{rho}
 \rho(a)=a^{-3}H(a)=a^{-3}\left(\frac{1}{2}\frac{p_{\phi}^2}{a^3}+a^3
 V(\phi)\right)
\end{equation}
with its potential $V(\phi)$ and momentum $p_{\phi}$. It can be
quantized by turning the momentum of $a$ into a derivative operator
acting on a wave function $\psi(a,\phi)$, resulting in the
Wheeler--DeWitt equation \cite{DeWitt,QCReview}
\begin{equation} \label{WdW}
 -\frac{1}{6}\ell_P^4 a^{-1}\frac{\partial}{\partial
  a}a^{-1}\frac{\partial}{\partial a} a\psi(a,\phi)=8\pi
  G\hat{H}(a)\psi(a,\phi)\,.
\end{equation}
Here, $\hat{H}(a)$ is the matter Hamiltonian acting on the scalar
dependence of $\psi$. It also depends on $a$ via the volume.
 
 In
this quantization we are not able to see any discrete picture or
other modification at small volume. In fact, this equation, though
quantized, cannot be shown to remove the singularity. One can see
that
 these problems are related to the fact that we just used
quantum
 mechanical techniques in going from the simple Friedmann
equation to
 the Wheeler--DeWitt equation. The same techniques cannot
be applied to
 more complicated systems, let alone the full
theory. It is then very
 likely that consistency conditions, which
would arise only in the
 complicated systems, are overlooked in the
quantization of the simple
 model. It would be more reliable if we
used a full quantum theory of
 gravity, such as loop quantum gravity,
and introduced the symmetries
 there. This is in fact possible
\cite{SymmRed}, and leads us to loop quantum
 cosmology where the
basic evolution equation for the isotropic case is \cite{IsoCosmo}
\begin{eqnarray} \label{Loop}
 -\frac{1}{2\sqrt{6}}\ell_P \left[
  \left(|n+2|^{3/2}-|n|^{3/2}\right)\psi_{n+1}(\phi)
 -2\left(|n+1|^{3/2}-|n-1|^{3/2}\right)\psi_{n}(\phi) \right.\\
 \left.+\left(|n|^{3/2}-|n-2|^{3/2}\right)\psi_{n-1}(\phi) \right] = 8\pi
  G\hat{H}(n)\psi_n(\phi)\,. \nonumber
\end{eqnarray}
It is immediately clear that the formulation is now discrete since we
have a
 difference equation in the integer $n$ replacing $a$ as a
label of the wave function, with the relation
$|n|=6a^2/l_P^2$. Another difference is that, unlike $a$, $n$ can
also
 take negative values, the sign corresponding to the orientation
of
 space. Most importantly, the equation is {\em non-singular}!
Starting from
 initial values for $\psi$ at large positive $n$, we
can evolve
 backwards up to and right through the classical
singularity at $n=0$
 \cite{Sing}. The evolution does not stop, and
we obtain a collapsing branch at
 negative $n$ preceding the
classical singularity. One must keep in
 mind, however, that the
classical space-time picture dissolves around
 $a=0$ and is replaced
by a discrete structure. A smooth transition, as sometimes presumed, is
impossible since smoothness would not even be defined.
 
\begin{figure}  
\vspace*{1.25cm}  
\begin{center}
\epsfig{figure=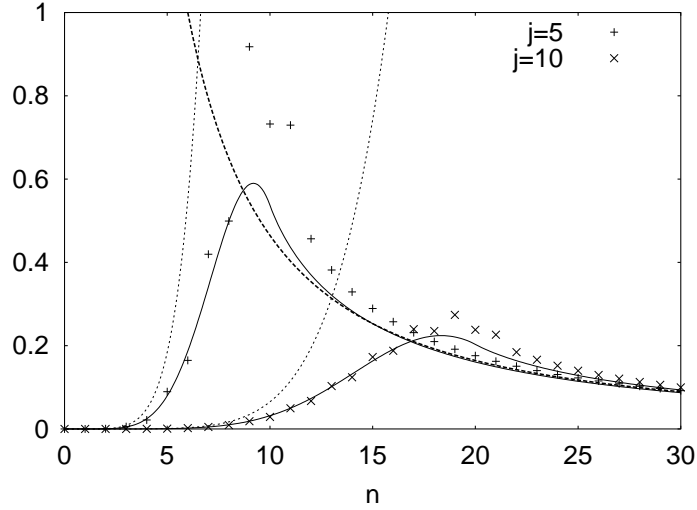,width=10cm}  
\end{center}
\vspace*{0.25cm}  
\caption{Eigenvalues of the density operator for two choices of the
  ambiguity parameter, compared to the classical expectation $a^{-3}$
  (thick, dashed). Also shown are continuous approximations to the
  discrete eigenvalues (solid), and small-volume approximations.  \label{Dens}
} 
\end{figure} 
 
 So far we only commented on the left hand side of Eq.~(\ref{Loop})
 which is obviously different from that of Eq.~(\ref{WdW}). The right
 hand side, however, is also changed because as matter
 Hamiltonian we
 have to use one which is related to that of the full
 theory. Since
 ultraviolet divergences are cut off there, also the
 divergence of
 $a^{-3}$ in the kinetic term of (\ref{rho}) is cut off
 at small
 scales, which has consequences regarding the evolution of the early
 universe. (In fact, some
 aspects of these modifications are already
 important for the removal
 of the singularity \cite{Sing}.)
 
\section{Inflation}

 Quantum gravity is expected to provide a cut-off for curvatures
which
 would otherwise diverge when a cosmological singularity is
approached. In the isotropic context, curvature components are
proportional to inverse powers of the scale factor $a$, for instance
the density $a^{-3}$ which also appears in the kinetic term of a
matter Hamiltonian (\ref{rho}). A natural cut-off is in fact realized
in loop
 quantum gravity, where quantization methods of the full
theory
 \cite{QSDV} imply a peak in the eigenvalues of an operator
quantizing
 $a^{-3}$ \cite{InvScale,Ambig}; see
Fig.~\ref{Dens}. Since this
 operator is not a basic one of the
quantum theory, it is subject to
 quantization ambiguities. In
particular the position of the peak
 changes when values
parameterizing the ambiguities are changed. This
 can easily be seen
in Fig.~\ref{Dens}, where the eigenvalues are
 plotted for two
choices of a parameter $j$ (a half-integer). Also the
 peak in the
eigenvalues (at a scale factor $a\sim \sqrt{j}\ell_P$) is
 obvious,
as well as the fact that for even smaller volume the
 eigenvalues
decrease rather than showing the classical
 divergence. This
demonstrates the expected curvature cut-off by
 quantum gravity
effects.

It is possible to approximate the discrete eigenvalues by a continuous
curve, which does not diverge at $a=0$,
\begin{equation} \label{ainvm}
 (a^{-3})^{(j)}=a^{-3} p(3a^2/j\ell_P^2)^{6}
\end{equation}
depending on the parameter $j$ (the approximation becomes better for
larger $j$, as can be seen in Fig.~\ref{Dens}). The function
\begin{eqnarray} \label{pq}
 p(q) &=& {\textstyle\frac{8}{77}}\, q^{1/4}\left[ 7\left(
 (q+1)^{11/4}-|q-1|^{11/4}\right)\right.\nonumber\\
 && - \left.11q \left( (q+1)^{7/4} -
 \sgn(q-1) |q-1|^{7/4}\right)\right]\;,
\end{eqnarray}
is derived from the quantum theory \cite{Ambig} (but also subject to
minor ambiguities) and provides the {\em interface to cosmological
phenomenology\/} in the following way: When the matter Hamiltonian is
derived from the quantum theory, the density $a^{-3}$ in its kinetic
part must show the cut-off. We can explicitly realize that by
replacing $a^{-3}$ with the modified $(a^{-3})^{(j)}$ for some
half-integer $j$. (Only the factor $a^{-3}$ in the kinetic term is
changed, not the pre-factor of the Hamiltonian in the density
(\ref{rho}), since the Hamiltonian is the primary object for the
quantization. Dividing by $a^3$ to obtain the density is done at the
classical level which cannot receive quantum modifications.)
In this way, we obtain the {\em effective Friedmann
equation}
\begin{equation} \label{effFriedmann}
 \left(\frac{\dot{a}}{a}\right)^2= \frac{16\pi}{3}G a^{-3}\left(\frac{1}{2}
 a^{-3} p(3a^2/j\ell_P^2)^6 p_{\phi}^2+ a^3V(\phi)\right)\,.
\end{equation}

\begin{figure}
\vspace*{1.25cm}  
\begin{center}
\epsfig{figure=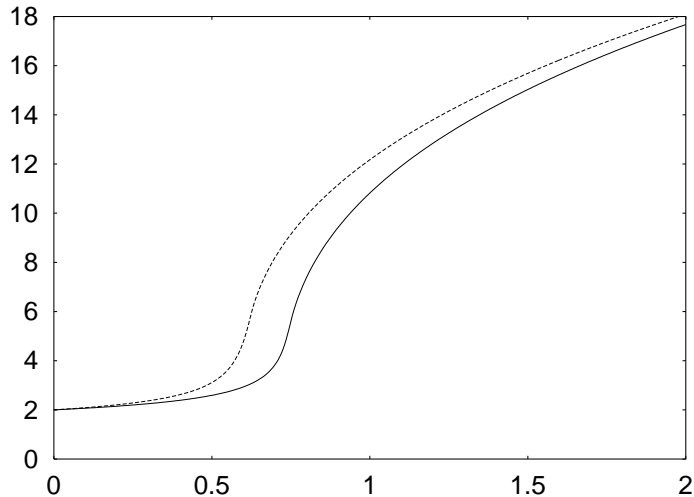,width=10cm}  
\end{center}
\vspace*{0.25cm}  
\caption{A numerical solution $a(t)$ of the effective Friedmann
  equation with vanishing potential (solid) and a small quadratic
  potential (dashed). The ambiguity parameter is $j=100$.
  \label{Infl}
} 
\end{figure} 

Since the right hand side now depends differently on $a$ for small $a$
compared to the classical behavior,
the dynamics is clearly modified. In particular, since the function
$p$ in (\ref{pq}) is increasing as a function of its argument when it
is small, the matter Hamiltonian at the right hand side is an
increasing function of the volume at small volume. Thermodynamically,
this implies {\em negative pressure\/} and therefore {\em inflation\/}
\cite{Inflation}. In fact, simple numerical solutions of the effective
Friedmann equation (\ref{effFriedmann}) clearly show an early phase of
accelerated expansion (Fig.~\ref{Infl}).

Thus, quantum geometry provides a new mechanism for inflation. It is a
consequence of a kinetic term modified by non-perturbative quantum
effects and is quite independent of the particular potential: even a
zero potential implies inflation. Furthermore, it is not necessary to
introduce an inflaton field, since any matter component will show the
modification and therefore lead to inflation via its kinetic term. The
details of the potential are, however, important for the observational
viability of an inflationary scenario.

Further possibilities for model building arise from the fact that matter
fields are driven away from their potential minima during quantum
geometry inflation because the usual friction term changes sign
\cite{Closed}: From the effective matter Hamiltonian
\[
 H^{\rm eff}(a)= \frac{1}{2}
 a^{-3} p(3a^2/j\ell_P^2)^6 p_{\phi}^2+ a^3V(\phi)
\]
at the right hand side of (\ref{effFriedmann}) we obtain the
Hamiltonian equations of motion
\[
 \dot{\phi}=\{\phi,H^{\rm eff}(a)\}= a^{-3} p(3a^2/j\ell_P^2)^6 p_{\phi}
\]
and
\[
 \dot{p}_{\phi}=\{p_{\phi},H^{\rm eff}(a)\}= -a^3 V'(\phi)
\]
for the scalar and its momentum. Both equations yield a second order
equation
\[
 \ddot{\phi}=p_{\phi}\frac{\md [a^{-3} p(3a^2/j\ell_P^2)^6]}{\md
 t}+a^{-3} p(3a^2/j\ell_P^2)^6 \dot{p}_{\phi}=
 a\frac{\md\log [a^{-3} p(3a^2/j\ell_P^2)^6]}{\md a}\, H\dot{\phi}-
 p(3a^2/j\ell_P^2)^6\, V'(\phi)
\]
for $\phi$. For large $a$, the function $p$ is close to one and we
obtain the usual friction term $-3H\dot{\phi}$ which dampens the
evolution of $\phi$. For small $a$, however, we noted repeatedly that
the effective density $a^{-3}p(3a^2/j\ell_P^2)^6$ increases as a
function of $a$. Thus, the derivative in the friction term has the
opposite sign and $\phi$ is driven up its potential.
This mechanism can be used to drive a subsequent phase of slow-roll
inflation, or may have consequences for structure generation or
reheating. These possibilities are currently being investigated.

\section{Conclusions} 
 
With new developments in quantum geometry, quantum gravity has become
a theory which can make concrete predictions about the very early
stages of the universe. Results include possible solutions of old
conceptual problems, as the singularity problem \cite{Sing} and the
problem of initial conditions \cite{DynIn}, and also new
phenomenological proposals
which can be confronted with cosmological observations
\cite{Inflation}. The models currently available are most likely too
simple, but more complicated ones with less symmetries (e.g.,
\cite{HomCosmo}) and more
realistic matter content are being developped. An advantage of the
formalism is that the relation between models and the full theory of
loop quantum gravity is known so that lessons learned for models can
be taken over to the full theory. In this way we will be able to
guide
 developments in quantum gravity by cosmological observations.

\acknowledgements{The author is grateful to Jean-Marc Virey for an
  invitation to the conference ``Where Cosmology and Fundamental
  Physics Meet,'' IUFM, Marseille where this work has been presented,
  and to Carlo Rovelli and Thomas
 Sch\"ucker for hospitality at CPT
  Marseille. This work was supported
 in part by NSF grant
  PHY00-90091 and the Eberly research funds of
 Penn State.  }

 
\vfill 
\end{document}